\documentclass[prb,twocolumn,superscriptaddress,aps]{revtex4-1}
\usepackage{amsmath}
\usepackage{color}
\usepackage{dcolumn}
\usepackage{bm}
\usepackage{slashed}
\usepackage[toc,page]{appendix}
\usepackage{float}
\usepackage{amssymb}

\usepackage[pdftex]{graphicx}

\def\gapp{\lower.35em\hbox{$\stackrel{\textstyle>}{\sim}$}}
\def\lapp{\lower.35em\hbox{$\stackrel{\textstyle<}{\sim}$}}

\begin{document}
\bibliographystyle{apsrev4-1}

\title{Non-abelian gauge fields and quadratic band touchings in molecular graphene}

\author{Fernando de Juan}
\affiliation{Materials Science Division, Lawrence Berkeley National Laboratories, Berkeley, CA
94720, USA}
\affiliation{Department of Physics, University of California, Berkeley, CA 94720, USA}
\date{\today}

\begin{abstract}
Dirac fermions in graphene can be subjected to non-abelian gauge fields by implementing certain
modulations of the carbon site potentials. Artificial graphene, engineered with a lattice of CO
molecules on top of the surface of Cu, offers an ideal arena to study their effects. In this work,
we show by symmetry arguments how the underlying CO lattice must be deformed to obtain these gauge
fields, and estimate their strength. We also discuss the fundamental differences between abelian
and non-abelian gauge fields from the Dirac electrons point of view, and show how a constant
(non-abelian) magnetic field gives rise to either a Landau level spectrum or a quadratic band
touching, depending on the gauge field that realizes it (a known feature of non-abelian gauge fields
known as the Wu-Yang ambiguity). We finally present the characteristic signatures of these effects
in the site-resolved density of states that can be directly measured in the current molecular
graphene experiment, and discuss prospects to realize the interaction induced broken symmetry states
of a quadratic touching in this system. 
\end{abstract}

\maketitle

\section{Introduction}

Condensed matter systems that host Dirac fermions as their electronic excitations have drawn a lot
of attention in recent years as they have become more and more experimentally accesible and
controlable, with graphene\cite{CGP09} and topological insulators\cite{HK10} being the most
prominent examples of such materials. 

A remarkable feature of Dirac fermions realized in graphene's
honeycomb lattice in particular is that one can further manipulate them externally by inducing
controlled strains in the sample, which couple to them as an effective gauge potential\cite{VKG10}.
This idea of strain engineering\cite{PC09} has led to many interesting
predictions\cite{GKG10,JCV11}, and is most spectacularly illustrated by the Landau level spectrum
recently observed\cite{LBM10} in scanning tunneling microscopy (STM). This system proved to be very
versatile, and in search for even better tunability several proposals were conceived to make
artificial versions of it\cite{PL09,GSP09,SGK11}. In a recent experimental breakthrough, a
realization of this type of systems, termed molecular graphene\cite{GMK12}, was built, which allows
for almost complete control of the electronic degrees of freedom within it. In this system, a
triangular lattice of CO molecules is assembled in the surface of bulk Cu, confining the surface
electrons to move in an effective hexagonal potential. In this way, effective Dirac fermions emerge
at the $K$ points of the superlattice potential, which can then be probed directly with an STM.

This system thus offers wide tunability to modify the electronic structure of the surface states by
distorting the CO lattice in any desired way, or by adding new atoms to the existing
structure. Indeed, several remarkable phenomena have already been demonstrated\cite{GMK12} beyond
the strain induced Landau levels, such as the opening of a gap by means of a Kekul\'e
distortion or the creation of an n-p-n junction. Other interesting proposals such as the
observation of fractional charge in a vortex \cite{HCM07,B12} or the synthesis of a quantum spin
Hall phase\cite{GGH12} should also be experimentally accesible. 

As noted in ref. \onlinecite{GGR12}, artificial graphene should be also ideal to explore the more
recent prediction that a full $SU(2)$ non-abelian gauge field is in fact realizable in this
system, and the strain induced one is just one component of it. Non-abelian gauge fields (of
singular nature) were known to emerge in graphene due to disclinations in the lattice\cite{GGV93},
but they can also be generated in a smooth fashion by modulating the on-site potential of
the carbon atoms in a certain way. As we will discuss, the effects of non-abelian gauge fields
can be very different from their abelian counterparts, and it is the purpose of this work to discuss
how to adapt the molecular graphene experiment to probe these differences. In particular, we
will show that a quadratic band touching can be generated with these fields, allowing a controlled
simulation of this band structure which is prone to many-body instabilities\cite{WAF10,MEM11,VJB12}.

In general, effective external gauge fields acting on a fermion system may have non-abelian
structure when the fermions have internal degrees of freedom, and the gauge field is a matrix
acting on this degree of freedom $\vec A_{ab}$ whose components need not commute. A typical
condensed matter example is spin and the spin-orbit interaction, which can be modeled as an SU(2)
gauge
field \cite{FS93,T08}, but there are many more examples \cite{WZ84,OBS05,RJO05,DGJ11}. A more recent
one is bilayer graphene \cite{SGG12}, where the two components of the SU(2) doublet correspond to
the wave functions in the two layers, and the interlayer interaction plays the role of the gauge
field. In the case of monolayer graphene, the SU(2) doublet is made with the valley degree of
freedom \cite{GGR12}. 

The non-abelian field strength is defined in terms of the covariant derivative $D_i = \partial_i
-iA_i$ as $F_{ij} = [D_i,D_j] = \partial_i A_j - \partial_j A_i - i[A_i,A_j]$,
which in two dimensions gives rise to a non-abelian magnetic field of the form
\begin{equation}
B^{\alpha} = \vec \partial \times \vec A^{\alpha} + \epsilon^{\alpha \beta \gamma}
 \vec A^{\beta} \times \vec A^{\gamma},\label{magnetic}
\end{equation}
where $\vec A_{ab} = \vec A^{\alpha} \Lambda^{\alpha}_{ab}$ with $\Lambda^{\alpha}_{ab}$ the
generators of SU(2), repeated indices are summed, and $\alpha=x,y,z$ (the indices $ab$ will be
implicit from now on). The last term in this expression arises because of the non-commutativity of
the field components and makes non-abelian gauge fields fundamentally different from their abelian
counterparts. In particular, it is responsible for a tricky feature of these gauge fields known as
the Wu-Yang ambiguity \cite{WY75}: the fact that one may have physically distinct gauge fields
(i.e. not gauge equivalent) with the same magnetic field. Indeed, consider these two simple
examples\cite{FSW97}. The first (type I) is $\vec A^{(3)} = B/2 (-y,x)$, $\vec A^{(1)}=\vec
A^{(2)}=0$,
which we recognize as the analog of the symmetric gauge for constant (abelian) magnetic field $B$,
in this case in the $z$ direction. The second (type II) is $\vec A^{(1)} = \sqrt{B/2}(1,0)$, $\vec
A^{(2)} = \sqrt{B/2}(0,1)$ and $\vec A^{(3)} = 0$, it also gives constant field $B_0$ due to the
second term in eq. (\ref{magnetic}), and it is not gauge related to type I. The magnetic field
alone is therefore not enough to distinguish these two cases\footnote{The resolution to this
apparent paradox
is simply that for non-abelian fields there are further independent gauge invariant quantities, such
as $D_iF_{jk}$, that distinguish among them.}, but we will see that the spectrum obtained for each
case is very different, and this is the physics that, as we will show, can be probed directly in
the molecular graphene experiment.
\begin{center}
\begin{table}
\begin{tabular}{|c||c|c|c|c|}
\hline
 & \multicolumn{2}{|c|}{Valley diagonal} & \multicolumn{2}{|c|}{Valley off-diagonal}  \\ \hline
Rep. & Symm. adapted & $\sigma_i \otimes \tau_j$ & Symm. adapted & 
 $\sigma_i \otimes \tau_j$ \\ \hline
$A_1$ & $\mathcal{I}$ & $\mathcal{I}$ & $\Lambda_x\Sigma_z$ & $\sigma_x \tau_x $ \\ \hline
$B_1$  & $\Lambda_z$ & $\tau_z$  & $\Lambda_y\Sigma_z$ & $\sigma_x \tau_y$  \\ \hline
$A_2$ & $\Sigma_z$ & $\sigma_z\tau_z$ & $\Lambda_x$ & $-\sigma_y\tau_y$  \\ \hline
$B_2$ & $\Sigma_z\Lambda_z$ & $\sigma_z$ & $\Lambda_y$ & $\sigma_y\tau_x$ \\ \hline
$E_{1x}$ & $\Sigma_x$ & $\sigma_x\tau_z$  &
$\Lambda_x\Sigma_y$ & $-\tau_y$ \\ \hline
$E_{1y}$ & $\Sigma_y$ & $\sigma_y$  &
$-\Lambda_x\Sigma_x$ & $\sigma_z\tau_x$ \\ \hline
$E_{2x}$ & $-\Lambda_z \Sigma_y$ & $-\sigma_y\tau_z$ &
$\Lambda_y\Sigma_x$ & $-\sigma_z\tau_y$ \\ \hline
$E_{2y}$ & $\Lambda_z\Sigma_x$ & $\sigma_x$ &
$\Lambda_y\Sigma_y$ & $\tau_x$ \\ \hline
\end{tabular}
\caption{Classification of basis matrices in the low energy theory around the
$K$,$K'$ points in graphene according to the representations of the symmetry group $C_{6v}$, and
their explicit realization in the basis $(\psi_{AK},\psi_{BK},\psi_{AK'},\psi_{BK'})$
(see ref. \onlinecite{B08} for details).}\label{tab}
\end{table}
\end{center}
\section{Symmetry analysis and microscopic calculation}

To realize an SU(2) gauge field in graphene, what we need is to externally apply certain on-site
potential patterns\cite{GGR12} to the carbon atoms of the honeycomb
lattice. It is not a priori clear, however, how this may be achieved in a molecular graphene
experiment, where the ``effective honeycomb lattice`` is engineered with the potential landscape
induced by a triangular array of CO molecules. In terms of an effective tight binding model, it is
natural to think that small distortions of this triangular lattice will produce potential changes in
the effective carbon sites, but what distortions will give rise to the correct potentials? And more
importantly, since these distortions may induce changes in the effective hopping as
well\cite{GMK12}, is it possible to modulate \emph{only} the on-site potential? 

To answer these questions, a symmetry approach to the problem appears better suited. The way
that external perturbations couple to the low energy degrees of freedom around a
high-symmetry point of the Brillouin Zone can be determined just by symmetry arguments. This
approach has been fruitfully employed in graphene to discuss the coupling of phonons, strains, or
electromagnetic fields \cite{M07,B08,WZ10,L12} and we now show how it can be
used to see the emergence of non-abelian gauge fields from small CO displacements the molecular
graphene.

In the half-filled honeycomb lattice, electrons close to the Fermi surface live near the
$K$ and $K'$ points, and are described by an effective spinor
$(\psi_{AK},\psi_{BK},\psi_{AK'},\psi_{BK'})$, where $A/B$ denotes the sublattice degree of
freedom. The effective Hamiltonian is conventionally written in the basis of the Pauli matrices
$\sigma_i \otimes \tau_j$, where $\sigma_i$ acts on the sublattice and $\tau_i$ on the valley
degrees of freedom, and $i=x,y,z$ (the identity in both sets is understood to be included as part
of the basis). 

To exploit the fact that the Hamiltonian must be a scalar under the symmetry group $C_{6v}$ of the
honeycomb lattice, one can relabel these basis matrices in terms of a new symmetry adapted set
$\Sigma_i$ and $\Lambda_i$ with the Pauli matrix algebra and well-defined transformation properties
under this group (technically, the group is $C_{6v}''$ because the unit cell has been
tripled to consider $K$ and $K'$ at the same time. We will refer to the labels under $C_{6v}$ for
simplicity, see ref. \onlinecite{B08} for details). The relation of these matrices to the original
ones and the representations according to which they transform are reproduced in table \ref{tab}. 

In this basis, the low energy Hamiltonian is simply written as ($v_F=1$)
\begin{equation}
H = \vec \Sigma \cdot \vec k,
\end{equation}
and in this form it is simple to see that the matrices $\Lambda_i$ commute with the
Hamiltonian and generate an SU(2) symmetry, which corresponds to rotations in the valley degree of
freedom.  

A gauge field is by definition a field that couples minimally in the form $k_i \rightarrow
k_i + A_i$, and in analogy with the usual electromagnetic field that couples as $H_{U(1)} =
\vec \Sigma \cdot \vec A$, one may introduce an SU(2) gauge field that couples as
\begin{figure}[h]
\begin{center}
\includegraphics[width=8.7cm]{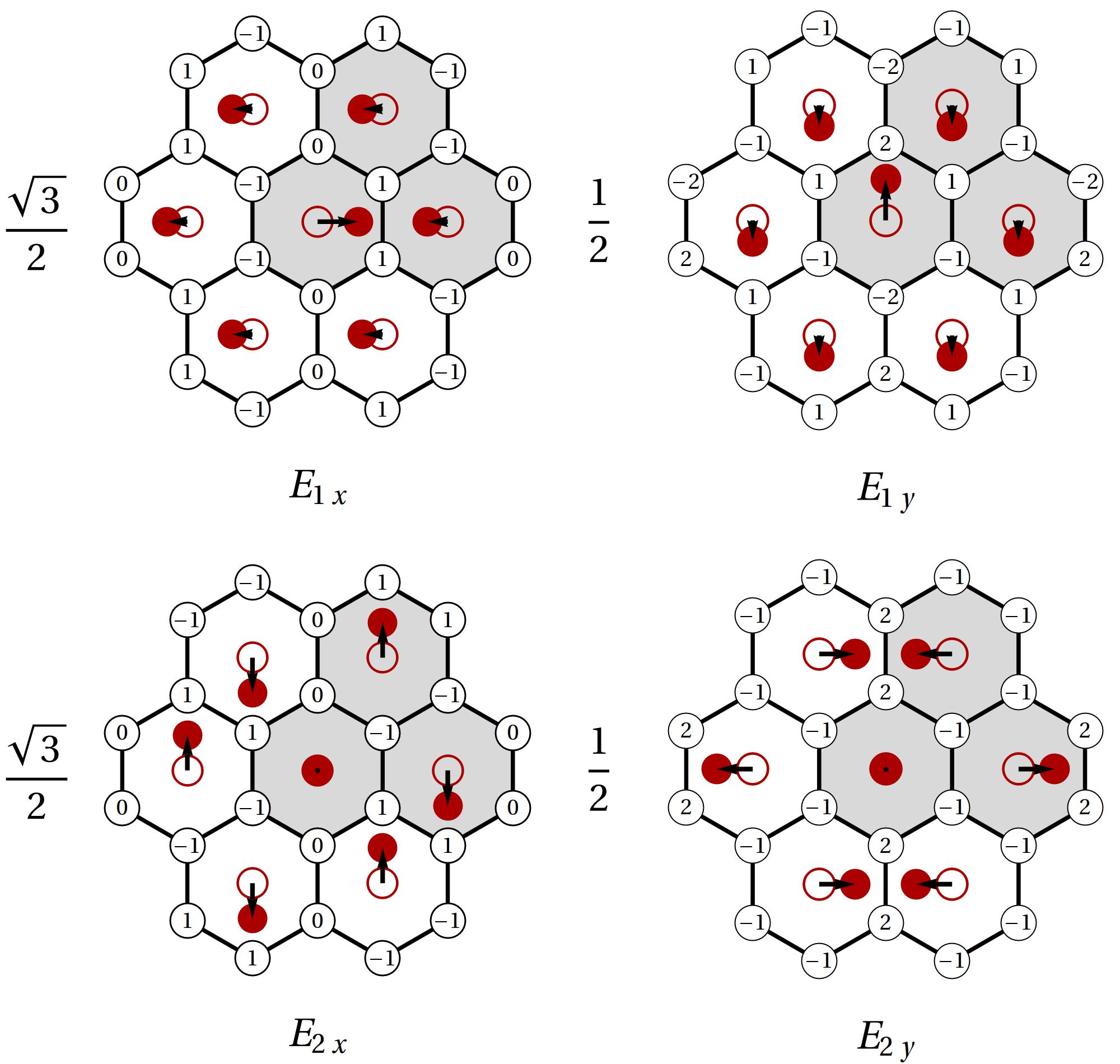}
\caption{(Color online) The four possible CO displacements with symmetries $E_1$ and $E_2$. The CO
molecules are represented in red, and the effective honeycomb lattice is shown in black. The unit
cell is shaded in gray, but more hexagons are shown to make the symmetry of the modes apparent. The
on-site potentials that match the symmetry labels are also shown in the effective carbon sites. Note
that the prefactors only refer to the potentials, not to the displacements. One may
think of these displacements as the K-point phonons of the triangular CO lattice.}\label{phonon}
\end{center}
\end{figure}
\begin{align}
H_{SU(2)} &= \vec \Sigma \cdot  \left( \Lambda_x \vec A^{(x)} +
\Lambda_y \vec A^{(y)} +\cdot \Lambda_z  \vec A^{(z)}\right),
\end{align}
which is a coupling allowed by symmetry if the gauge fields $\vec A^{\alpha}$ have their origin in a
microscopic perturbation with the same symmetry as the matrix that accompanies them. 

The power of the symmetry analysis is thus that one can now say what type of perturbations
correspond to each term only by inspection of table \ref{tab}. Perturbations in the first column
have the periodicity of the unit cell, while those in the second column have the periodicity of a
tripled unit cell (because of intervalley mixing). Moreover, within nearest neighbour tight
binding (TB), those
perturbations diagonal in sublattice ($\propto \sigma_0$ or $\sigma_z$) correspond to potential
modulations, while those off diagonal correspond to hopping modulations. With this criterion, the
gauge field $\vec A^{(z)}$ is readily identified as the usual strain-induced gauge field. The gauge
field components $\vec A^{(x)}$ and $\vec A^{(y)}$ correspond, respectively, to the valley
mixing $E_1$ and $E_2$ potential perturbations defined in ref. \onlinecite{B08} (which are labeled
$G'$ under $C_{6v}''$). Their corresponding potentials are depicted in fig. \ref{phonon} in the
effective carbon sites. 
\begin{figure}[h]
\begin{center}
\includegraphics[width=8.7cm]{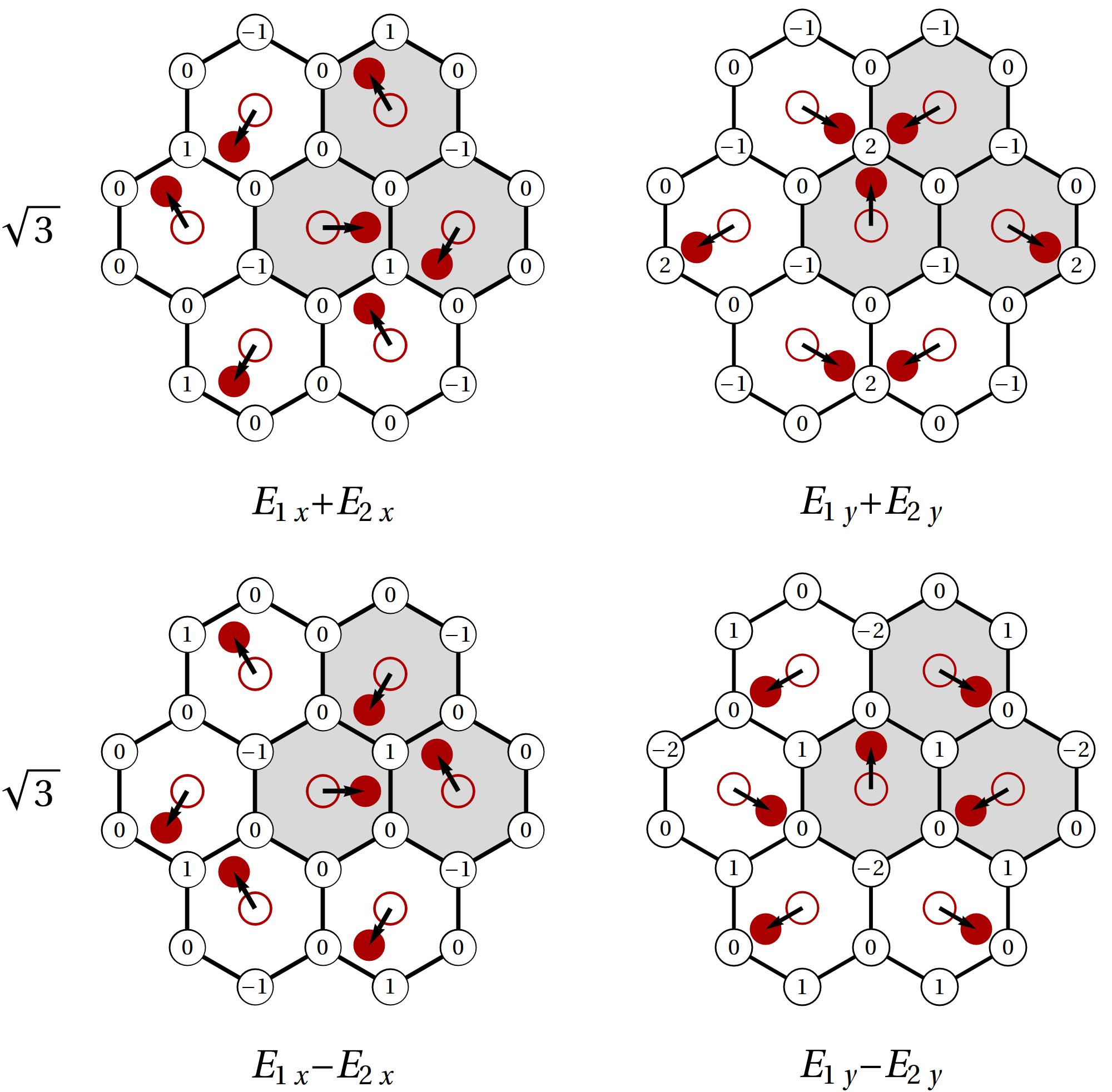}
\caption{(Color online) With the same conventions that fig. 1, combinations of CO displacements that
produce a quadratic band touching. Again, note that the prefactors only refer to the potentials, not
to the displacements.}\label{cuad}
\end{center}
\end{figure}
In real graphene this type of potential perturbation is the one induced by phonons like the LO/LA
phonon at the K point \cite{Falko}, or the ZO/ZA phonon at the K point in the presence of a
perpendicular electric field, and it is known that it can also be produced by a particular substrate
\cite{Pankratov}.

For molecular graphene, this analysis immediately allows to find the CO displacements that will
induce these potential modulations. These should be displacements with a tripled unit cell and the
appropriate symmetry labels, and in fact may be simply interpreted as the $E_1$ and $E_2$ phonons of
the triangular CO lattice at the $K$ point. These displacements are, for three consecutive CO
molecules
\begin{align}\nonumber
\vec r_{CO,E_{1x}} &=  \left\{  (1,0) \; , \; (-\tfrac{1}{2},0)  \; , \; (-\tfrac{1}{2},0)
\right\}, \\ \nonumber
\vec r_{CO,E_{1y}} &= \left\{ (0,1) \; , \; (0,-\tfrac{1}{2})\; , \; (0,-\tfrac{1}{2})\right\}, 
\end{align}
\begin{align}
\nonumber
\vec r_{CO,E_{2x}} &= \left\{ (0,0) \; , \; (0,-\tfrac{\sqrt{3}}{2}) \; , \;(0,\tfrac{\sqrt{3}}{2})
\right\} ,\\ \nonumber
\vec r_{CO,E_{2x}} &= \left\{ (0,0) \; , \;(\tfrac{\sqrt{3}}{2},0)\; , \;
(-\tfrac{\sqrt{3}}{2},0)\right\} ,
\end{align}
and are also shown in fig \ref{phonon}. Indeed, within a
TB model one can parametrize the change in on-site potential with displacement as
\begin{equation}
V_i = V' \sum_j \Delta \vec r_{j,CO} \cdot \vec \delta_{ij}, \label{pot}
\end{equation}
for a carbon site $i$ with $j$ CO neighbours at equilibrium distances $\delta_{ij}$ from it, and
verify that the potentials shown in fig. \ref{phonon} are given by eq. (\ref{pot}). The constant
$V' \equiv \partial V /\partial a$ parametrizes the change in on-site potential with distance, and
may be estimated by realizing that this physical mechanism is responsible for the scalar potential
$\phi$ in the continuum Dirac equation. A comparison with the p-n junction experiment yields $V'
\approx 22$ $\text{meV}/\text{\AA}$ (see appendix), which is very similar to $\partial t /\partial
a = \beta t/a \approx 20$ $\text{meV}/\text{\AA}$. This is also consistent with the fact that in
real graphene the analog of $V'$ for carbon displacements\cite{Falko} is of the same order as
$\partial t /\partial a$. 

Finally, the symmetry analysis also reveals that close to the Dirac point, the desired CO
displacements do not introduce any other change in the effective theory other than the
$\vec A^{(x)},\vec A^{(y)}$ gauge fields. In particular, while these displacements may induce
nearest neighbour hopping changes, these cannot appear in the low energy theory because there are no
intervalley matrices in the $E_1$ or $E_2$ representations that are sublattice off-diagonal. This
hopping changes thus have no effect in the low energy properties and we will not consider them
in what follows. Changes in the next nearest neighbour hopping $t'$ due this displacements are
small and need not be considered.

To obtain the gauge field from a microscopic calculation, one may substitute eq. (\ref{pot}) in the
effective tight binding model
\begin{equation}
H = -t\sum_{\left<i,j\right>} c^{\dagger}_i c_j -t'\sum_{\left<\left<i,j\right>\right>}
c^{\dagger}_i c_j + \sum_i V_i c^{\dagger}_i c_i. \label{TBH}
\end{equation}
The potential modulation (depicted in fig. \ref{phonon}) that gives rise to the non-abelian gauge
fields is \cite{GGR12}
\begin{align}
V(\vec x) = \frac{3}{2} V' \left[ u_{E_{2y}} \cos \vec K \vec x + u_{E_{1x}} \sin \vec K \vec x
\right. \\
 +\left. \frac{2}{\sqrt{3}} \sin \vec G \vec x \left(u_{E_{1y}} \cos \vec K \vec
x+u_{E_{2x}} \sin \vec K \vec x \right)  \right], \nonumber
\end{align}
with $\vec K=(4\pi/3\sqrt{3},0)$ a vector joining the two dirac points and $\vec G=(0,-4\pi/3)$ a
reciprocal lattice vector. To project this perturbations into the Dirac points one performs the sum 
\begin{equation}
H = \sum_i V_i c^{\dagger}_i c_i = \sum_{\vec x} V(\vec x) c^{A,\dagger}_{\vec x}
c^A_{\vec x} + V(\vec x + \vec \delta_1) c^{B,\dagger}_{\vec x} c^B_{\vec x}, 
\end{equation}
with $\vec x = n \vec a_1 + m\vec a_2$ the lattice positions, $\vec \delta_1=a(0,1)$ a nearest
neighbour vector, and 
\begin{align}
c^A_x = e^{i \vec K \vec x} c^A_K + e^{-i \vec K \vec x}c^A_{K'}, \\
c^B_x = e^{i \vec K \vec x} c^B_K + e^{-i \vec K \vec x}c^B_{K'}.
\end{align}
This sum gives exactly the matrices dictated by symmetry
\begin{equation}
H = \frac{3}{4} V' (-\tau_2  u_{E_{1x}} + \tau_1 \sigma_3 u_{E_{1y}}-\tau_2 \sigma_3
u_{E_{2x}} + \tau_1 u_{E_{2y}}),
\end{equation}
so that the final formula relating the effective gauge fields to CO displacements
is
\begin{align}
\vec A^{(1)} &=  \frac{3}{4} V' (-u_{E_{1y}},u_{E_{1x}}), \\
\vec A^{(2)} &=  \frac{3}{4} V' (u_{E_{2x}},u_{E_{2y}}). \label{dispcorresp}
\end{align}

\section{Physical effects}

\begin{figure}[h]
\begin{center}
\includegraphics[width=7cm]{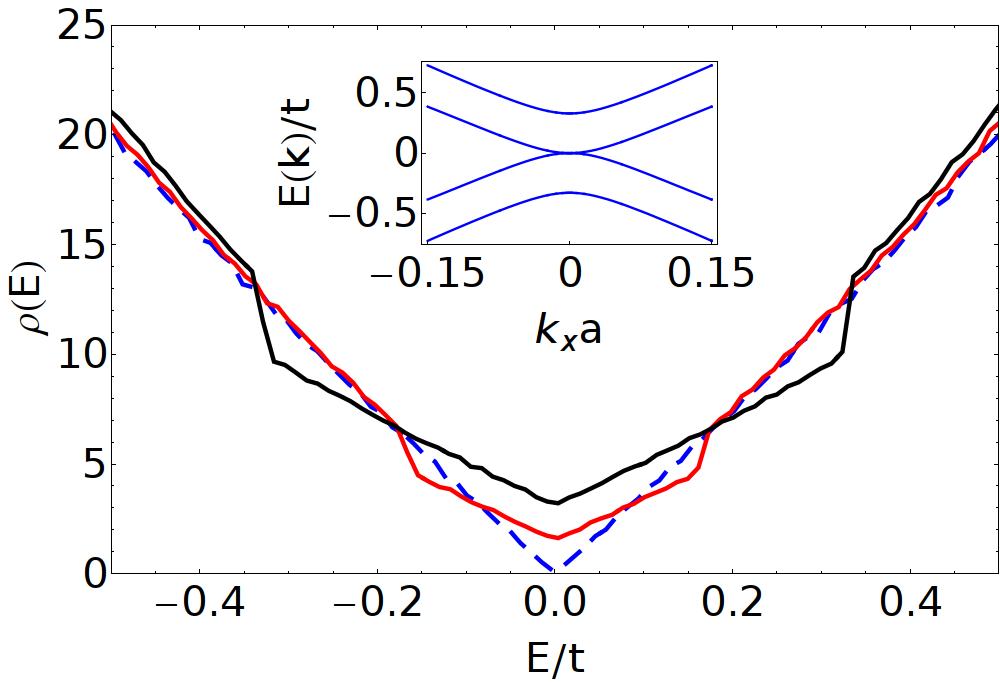}
\caption{(Color online) Total density of states for any type II gauge field of strength $u=0.5
\text{\AA}$ (red line) and $u=1\text{\AA}$ (black line), for $t=90$ meV and $t'=0$. The unperturbed
LDOS is shown as a dashed blue line for comparison. Inset: Band structure of the system for $u=1
\text{\AA}$. Note the similarity with bilayer graphene.}\label{total}
\end{center}
\end{figure}
As described in the introduction, we now consider two gauge field configurations that are not
related by a gauge transformation, but whose magnetic field is the same, and consider how they
should be seen in a local density of states (LDOS) measurement. Consider the type I gauge field,
with a magnetic field
pointing in a general direction $b^{\alpha}$ in $SU(2)$ space, $A_i^{\alpha} = b^{\alpha}
B/2(y,-x)$. When $b^{\alpha}=(0,0,1)$ we have the usual strain induced gauge field. The case
$b^{\alpha}=(0,1,0)$ was discussed in ref. \onlinecite{GGR12}. In general, by a constant $SU(2)$
rotation of the Hamiltonian, it is not difficult to see that for any $b^{\alpha}$ the spectrum is
still given by Landau levels $E_n = \sqrt{2Bn}$. The only difference appears in the
wavefunctions, because the sublattice polarization turns out to be given by the projection of
$b^{\alpha}$ onto the $z$ axis. For strain induced fields it is maximum, but for potential induced
ones the density of states is in fact constant across the unit cell. One can estimate the magnetic
field induced in the molecular graphene experiment with these gauge fields as follows. Take
$u_{E_{1x}} = u_{max}/L * x$, with $L$ the radius of the (approximately circular) sample and
$u_{max}$ the maximum displacement (at $x=L$). The magnetic field is (recovering all units)
\begin{equation}
B^{(x)} = \frac{\hbar/e}{\hbar v_F} \frac{3V'}{4} \frac{u_{max}}{L}.  
\end{equation}
With $\hbar/e = 6.5 \cdot 10^4 T\text{\AA}^2$ and taking $u_{max} = 0.1 a$ and
$\sqrt{3}a/L \approx 1/10$ and a Fermi velocity\cite{GMK12} $\hbar v_F \approx 1.5
\text{eV}\text{\AA}$ we obtain $B \approx 3.75 T$, which is not very large compared to the strain
induced one that is typically achieved. 

\begin{figure}[h]
\begin{center}
\includegraphics[width=4.25cm]{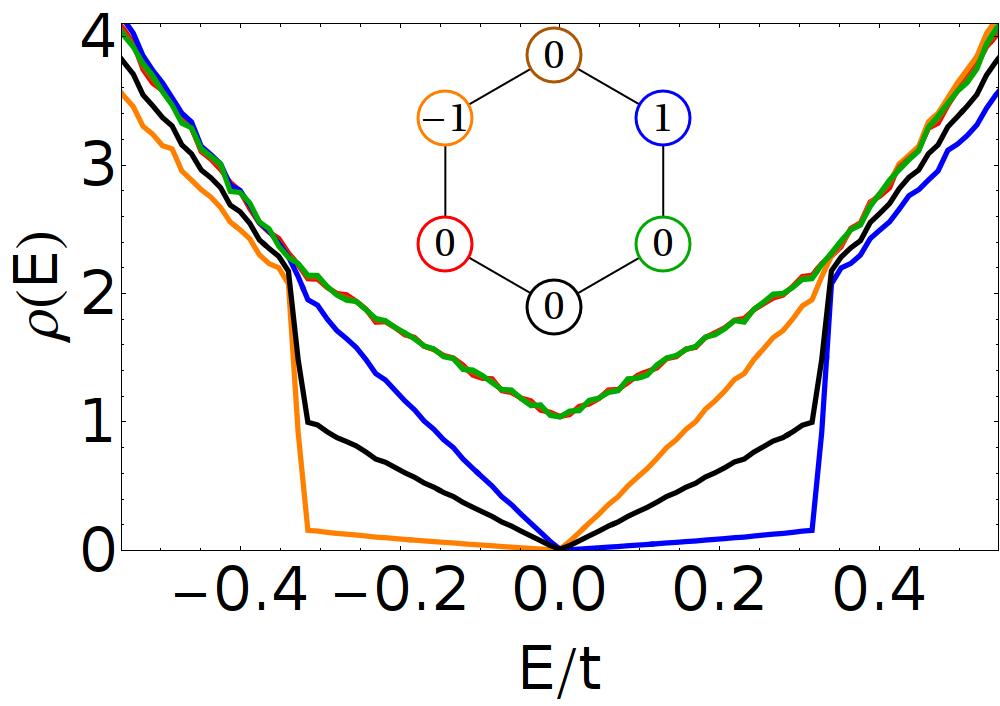}
\includegraphics[width=4.25cm]{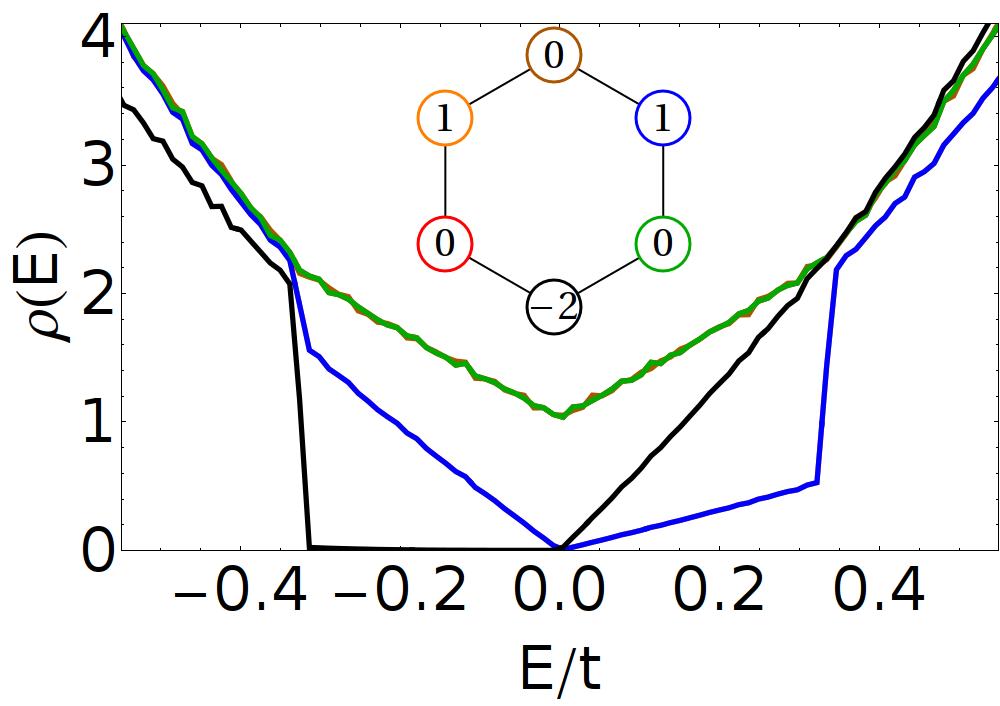}
\includegraphics[width=4.25cm]{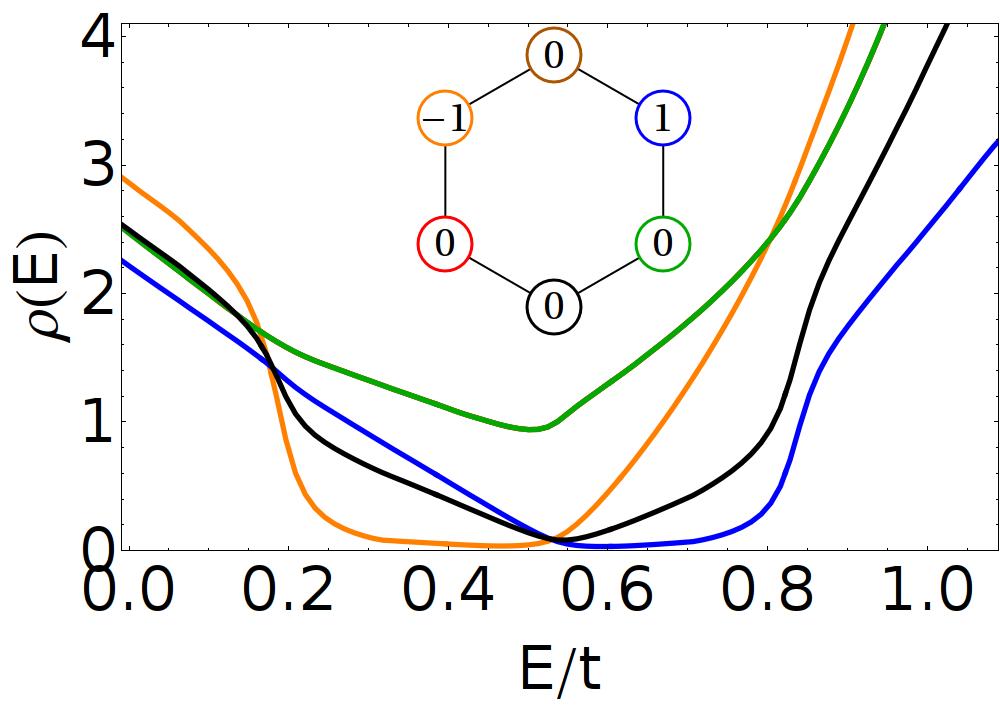}
\includegraphics[width=4.25cm]{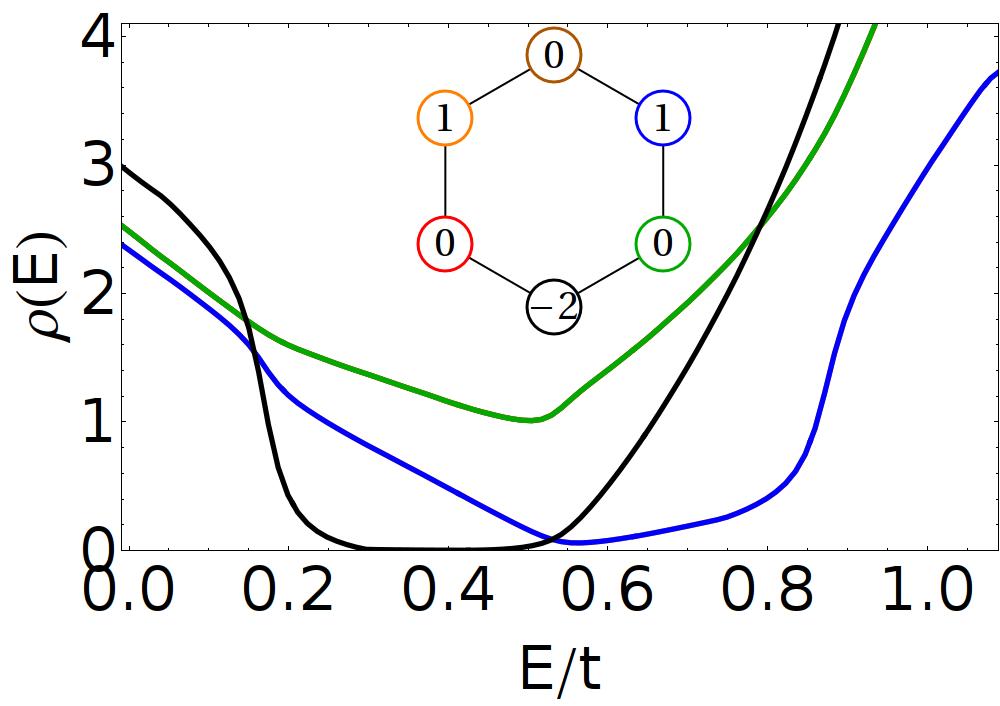}
\caption{(color online) LDOS as a function of energy for two different gauge fields. Top left:
$E_{1x}-E_{2x}$, with strength u=1 $\text{\AA}$, $t=90$ meV and $t'=0$. Top right: $E_{1y}-E_{2y}$,
same parameters. Bottom plots are the same but with $t'=0.18t$ and a Lorentzian broadening of
$\Sigma =0.2 t$. The insets show the corresponding on-site potential and the color code for the
different sites within the unit cell. Note that the missing lines in the plots overlap with the line
shown that has the same on-site potential.}\label{site}
\end{center}
\end{figure}

The type II gauge field has better prospects to be experimentally accessible. Keeping $\vec
A^{(3)}=0$, there are in fact four possible choices of constant gauge fields that give constant
magnetic field, given by
\begin{align}
\vec A^{1} = \sqrt{B/2}(1,0) & &\vec A^{2} = \sqrt{B/2} (0,\pm1), 
\end{align}
which, by eq. (\ref{dispcorresp}), is produced with the displacements $E_{1y}\pm E_{2y}$, and 
\begin{align}
\vec A^{1} =  \sqrt{B/2}(0,1) & &\vec A^{2} =  \sqrt{B/2}(\pm1,0),
\end{align}
which is produced with the displacements $E_{1x}\pm E_{2x}$. These displacements and their on-site
potentials are depicted in fig. \ref{cuad}. The magnetic field is given by $B = 9/8(V' u/v_F)^2$
with $u= u_{E_{1i}} = \pm u_{E_{2i}}$ representing the modulus of the displacements in fig.
\ref{cuad}. It is interesting to note that the estimate for $B$ in this case for $u= 0.1a$
is $B \approx 16 T$. 

The Dirac Hamiltonian in the presence of these gauge fields is formally analogous to that of
bilayer graphene (for a single valley), with the role of the layer played by the valley
here\footnote{Or to the Dirac Hamiltonian in the presence of Rashba spin-orbit coupling. Not
surprisingly these Hamiltonians are described in terms of SU(2) gauge fields too.}, and an
effective interlayer coupling $\gamma = \sqrt{2B} $. The spectrum of
these Hamiltonians is well known to be a quadratic band touching, with two extra parabolic bands at
higher energies. 

Considering first the case $t'=0$, the density of states (DOS) of this system is finite at the
touching point $E_D=0$,
and has a jump at $\pm v_F\sqrt{2B}= 3/2V' u$, as depicted in fig. \ref{total}. Considering a
displacement $u = 0.1a = 1\text{\AA}$, the kink in the LDOS should appear at $\pm \text{30 meV}$,
which should be easily observable. The precise location of this jump should serve as an
independent estimate of the parameter $V'$. The main effect of a finite $t'$ is to shift $E_D$ to a
higher value, as we will see below.

Moreover, this type of gauge field shows more complicated local density of states across the
enlarged unit cell. In fig. \ref{site} we show the LDOS for the cases $E_{1x}-E_{2x}$ and
$E_{1y}-E_{2y}$. The other two combinations are obtained by mirror symmetry. For $t'=0$, we observe
different local gaps for different sites, and finite LDOS at $E=0$. For more faithful comparison
with the experiment, we have also plotted the LDOS for $t'=0.18t$, and with a Lorentzian broadening
of $\Sigma=0.2t$. We observe the main effect of a shift in $E_D$, as well as some
electron-hole assymmetry, but the main features that characterize the non-abelian gauge field
remain.
The identification of these features in an STM measurement would represent a demonstration of the
presence of the type II constant non-abelian gauge field. 

\section{Discussion}

In this work we have shown, by means of a symmetry analysis, how non-abelian gauge fields may be
implemented in molecular graphene, and what their experimental signatures should be in the LDOS. For
type I gauge fields of constant magnetic field, we have shown that because of the different
microscopic origin of gauge fields $A^{(3)}$ (hopping change) and $A^{(1,2)}$ (potential change),
the magnetic field that one gets in the second
case is relatively smaller. While this may make the Landau level spectrum more difficult to
observe, the presence of this type of field could also be readily detected, for example, in a
quantum interference experiment in the weak field limit \cite{JCV11}. 

We have also shown that type II constant non-abelian gauge fields generate a quadratic band touching
analogous to bilayer graphene. Because of the enhanced DOS at the Fermi level, the electron-electron
interaction is known to drive this system to a broken symmetry state whose precise characteristics
are still controversial \cite{WAF10,MEM11,VJB12}. In the current molecular graphene
experiment, the Coulomb interaction is screened by the metallic bulk, leaving residual Hubbard
interactions estimated to be $U \sim 0.5t \sim 50$ meV (see Supplementary Material of ref.
\onlinecite{GMK12}). While an ideal quadratic touching is unstable to infinitesimal short range
interactions, the current broadening due to bulk tunneling ($\sim 0.2t$) is perhaps too large and
may challenge the observation of the interaction induced transition. Both bulk tunneling and
screening could be reduced by performing future experiments in bulk insulators with metallic
surfaces (such as the recently discovered topological insulators\cite{HK10}) which may eventually
allow to study the fate of the many body state with a tunable analog of the interlayer hopping.
Incidentally, it is also interesting to note that this instability can be interpreted as a
non-abelian magnetic catalysis, where an infinitesimal field drives chiral symmetry breaking
\cite{GHS98}. Furthermore, the controlled simulation of these non-abelian gauge fields, when made
position dependent, may be used to study the generation of zero-energy flat bands, as those observed
in the twisted bilayer system \cite{SGG12}, or the physics of topological defects in
the gauge field\cite{GGR12}.

In summary, the molecular graphene experiment has great potential to observe many interesting
phenomena related to non-abelian gauge fields with an unprecedent tunability, and which, as we have
shown, should be realizable in the current experimental samples. 

\section{Acknowledgements}

I would like to thank D. Rastawiki, V. Juricic, H. Ochoa, A. G. Grushin and H. Manoharan for useful
discussions. Funding from the ``Programa Nacional de Movilidad de Recursos Humanos" (Spanish MECD)
is acknowledged.

\section{Appendix}

\subsection{Estimate of V'}

The parameter $V'$ describes the change of on-site potential due to the displacements of
neighbouring $CO$ molecules. As such, it is featured both in the non abelian gauge fields (which
come from ''optical`` displacements) and in the strain-induced scalar potential $\phi$ (which comes
from ''acoustical`` displacements). The scalar potential $\phi$ also has a contribution from NNN
hopping change $ \partial t' /\partial a$ but it is much smaller and will be neglected.

To see this, consider an isotropic expansion of the $CO$ lattice. For every carbon site $i$, the
induced potential is given by eq. (\ref{pot}). Because displacement is smooth we may write 
\begin{align}
V_{\vec x} =& V' \sum_m \delta_m^i r^i_{x+\delta_m,CO} \approx V' \sum_m \delta_m^i \frac{\delta_m^j
\partial^j r^i_{x,CO}}{a} \nonumber \\
=& \frac{3a}{2} V' (u_{xx}+u_{yy}),
\end{align}
and plugging directly into the TB Hamiltonian eq. (\ref{TBH}), we obtain that $\phi = 3V'a/2(u_{xx}
+ u_{yy})$. Now consider the p-n-p juntion experiment in ref. \onlinecite{GMK12}. The middle region
is strained from $d=17.8$ to $d=20.4$ so $u_{xx} = u_{yy} = 0.14$. The change in scalar potential
$\Delta \phi$ is 95 meV, so we obtain ($d= \sqrt{3} a$)
\begin{equation}
V' = \frac{95 \text{meV}}{0.14 \sqrt{3} \; 17.8 \text{\AA}} = 22 \; \text{meV}/\text{\AA}.
\end{equation}
A different estimate can be obtained from the nearly free electron model considered in ref
\onlinecite{GMK12} (supp. mat.), where the scalar potential is
\begin{equation}
H = \frac{8 \pi^2}{9 d^2 m} (u_{xx}+u_{yy}) = \frac{3a}{2} V' (u_{xx}+u_{yy}),
\end{equation}
which gives $V' = 24 \; \text{meV}/\text{\AA}$

\subsection{Symmetry of hopping perturbations}

\begin{figure}[h]
\begin{center}
\includegraphics[width=9cm]{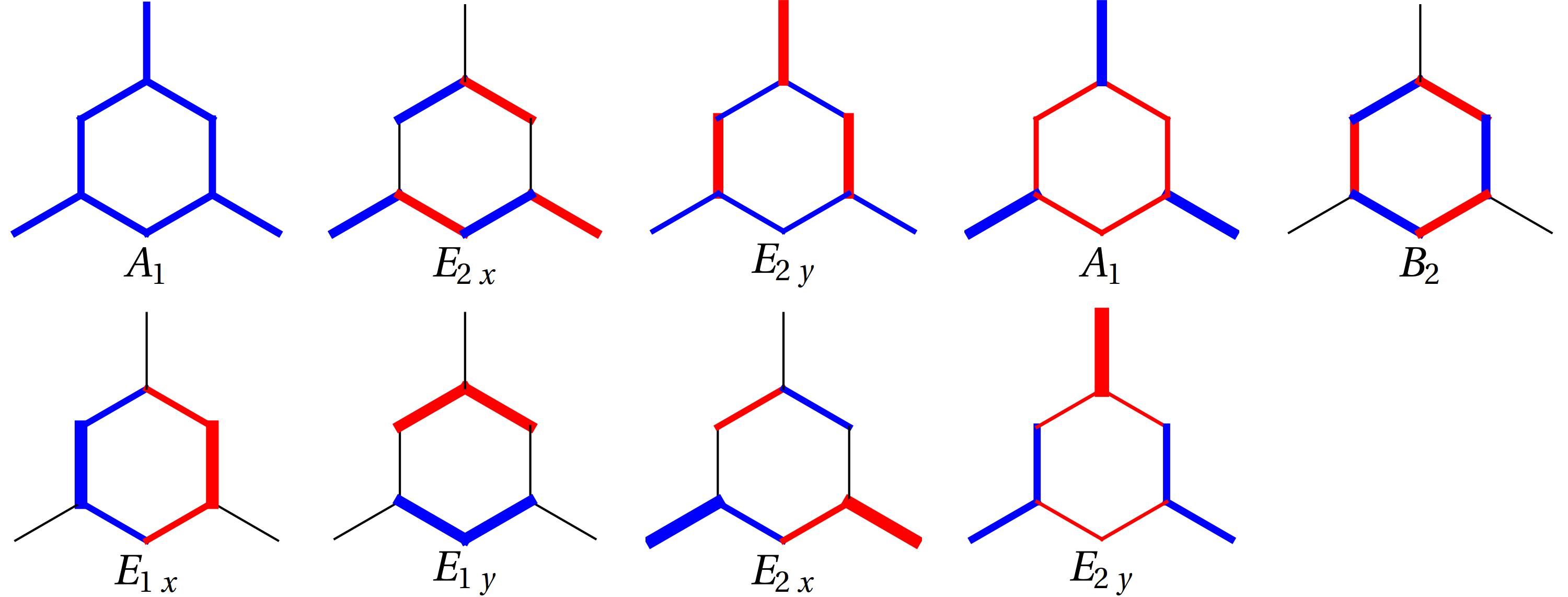}
\caption{(Color online) The 9 independent hopping patterns and their symmetry labels. Blue means
positive and red negative, and black lines represent no change in the hopping.  Hopping
modulations of the corresponding symmetry may also be induced by these displacements (red is
negative hopping and blue is positive. However, these particular patterns have no effect in the low
energy theory: they do not affect the quadratic touching and the LDOS predictions around
$E=E_D$.}\label{9hop}
\end{center}
\end{figure}

There are 9 independent hoppings in the tripled unit cell, which can be decomposed into
combinations that have well defined transformation properties under the symmetries of the lattice.
The 9 combinations and their symmetry labels are shown in fig. \ref{9hop}. In the first row, one
may identify the constant hopping ($A_1$), the $E_2$ pattern that gives rise to the usual gauge
field, and the Kekulé distortions (any of the three domains can be obtained from these). 

The four combinations in the second row are $E_1$ and $E_2$ (and form the representation $G'$ when
the enlarged group $C_{6v}''$ is considered). These hopping patterns are produced by the same
$CO$ displacements that give the non-abelian gauge fields through charge modulation. In the main
part of the text we claimed that these hopping distortions cannot couple to the low energy theory
around the $K$ point. The reason is simply that there is no valley off-diagonal $E_1$ or $E_2$
matrix in the low energy theory whose microscopic origin is a hopping change. This can be seen
directly by inspection of table \ref{tab}, where the valley mixing $E_1$ and $E_2$ matrices are all
diagonal in sublattice.  The hopping modulations will only appear in the low-energy theory if terms
with higher order in momentum are considered. If one is interested in the whole band structure and
not just low energies, these distortions in the hopping should be included by changing the NN
hopping in the usual manner. 

\bibliography{nonabelian}

\begin{thebibliography}{10}%
\makeatletter
\providecommand \@ifxundefined [1]{%
 \ifx #1\undefined \expandafter \@firstoftwo
 \else \expandafter \@secondoftwo
\fi
}%
\providecommand \@ifnum [1]{%
 \ifnum #1\expandafter \@firstoftwo
 \else \expandafter \@secondoftwo
\fi
}%
\providecommand \enquote [1]{``#1''}%
\providecommand \bibnamefont  [1]{#1}%
\providecommand \bibfnamefont [1]{#1}%
\providecommand \citenamefont [1]{#1}%
\providecommand\href[0]{\@sanitize\@href}%
\providecommand\@href[1]{\endgroup\@@startlink{#1}\endgroup\@@href}%
\providecommand\@@href[1]{#1\@@endlink}%
\providecommand \@sanitize [0]{\begingroup\catcode`\&12\catcode`\#12\relax}%
\@ifxundefined \pdfoutput {\@firstoftwo}{%
 \@ifnum{\z@=\pdfoutput}{\@firstoftwo}{\@secondoftwo}%
}{%
 \providecommand\@@startlink[1]{\leavevmode\special{html:<a href="#1">}}%
 \providecommand\@@endlink[0]{\special{html:</a>}}%
}{%
 \providecommand\@@startlink[1]{%
  \leavevmode
  \pdfstartlink
   attr{/Border[0 0 1 ]/H/I/C[0 1 1]}%
   user{/Subtype/Link/A<</Type/Action/S/URI/URI(#1)>>}%
  \relax
 }%
 \providecommand\@@endlink[0]{\pdfendlink}%
}%
\providecommand \url  [0]{\begingroup\@sanitize \@url }%
\providecommand \@url [1]{\endgroup\@href {#1}{\urlprefix}}%
\providecommand \urlprefix [0]{URL }%
\providecommand \Eprint[0]{\href }%
\@ifxundefined \urlstyle {%
  \providecommand \doi [1]{doi:\discretionary{}{}{}#1}%
}{%
  \providecommand \doi [0]{doi:\discretionary{}{}{}\begingroup
  \urlstyle{rm}\Url }%
}%
\providecommand \doibase [0]{http://dx.doi.org/}%
\providecommand \Doi[1]{\href{\doibase#1}}%
\providecommand \bibAnnote [3]{%
  \BibitemShut{#1}%
  \begin{quotation}\noindent
    \textsc{Key:}\ #2\\\textsc{Annotation:}\ #3%
  \end{quotation}%
}%
\providecommand \bibAnnoteFile [2]{%
  \IfFileExists{#2}{\bibAnnote {#1} {#2} {\input{#2}}}{}%
}%
\providecommand \typeout [0]{\immediate \write \m@ne }%
\providecommand \selectlanguage [0]{\@gobble}%
\providecommand \bibinfo [0]{\@secondoftwo}%
\providecommand \bibfield [0]{\@secondoftwo}%
\providecommand \translation [1]{[#1]}%
\providecommand \BibitemOpen[0]{}%
\providecommand \bibitemStop [0]{}%
\providecommand \bibitemNoStop [0]{.\EOS\space}%
\providecommand \EOS [0]{\spacefactor3000\relax}%
\providecommand \BibitemShut [1]{\csname bibitem#1\endcsname}%
\bibitem{CGP09}%
  \BibitemOpen
  \bibfield{author}{%
  \bibinfo {author} {\bibfnamefont{A.~H.}\ \bibnamefont{Castro~Neto}}, \bibinfo
  {author} {\bibfnamefont{F.}~\bibnamefont{Guinea}}, \bibinfo {author}
  {\bibfnamefont{N.~M.~R.}\ \bibnamefont{Peres}}, \bibinfo {author}
  {\bibfnamefont{K.~S.}\ \bibnamefont{Novoselov}},\ and\ \bibinfo {author}
  {\bibfnamefont{A.~K.}\ \bibnamefont{Geim}},\ }%
  \bibfield{journal}{%
  \Doi{10.1103/RevModPhys.81.109}{\bibinfo {journal} {Rev. Mod. Phys.}}\ }%
  \textbf{\bibinfo {volume} {81}},\ \bibinfo {pages} {109} (\bibinfo {year}
  {2009})%
  \bibAnnoteFile{NoStop}{CGP09}%
\bibitem{HK10}%
  \BibitemOpen
  \bibfield{author}{%
  \bibinfo {author} {\bibfnamefont{M.~Z.}\ \bibnamefont{Hasan}}\ and\ \bibinfo
  {author} {\bibfnamefont{C.~L.}\ \bibnamefont{Kane}},\ }%
  \bibfield{journal}{%
  \Doi{10.1103/RevModPhys.82.3045}{\bibinfo {journal} {Rev. Mod. Phys.}}\ }%
  \textbf{\bibinfo {volume} {82}},\ \bibinfo {pages} {3045} (\bibinfo {month}
  {Nov}\ \bibinfo {year} {2010})%
  \bibAnnoteFile{NoStop}{HK10}%
\bibitem{VKG10}%
  \BibitemOpen
  \bibfield{author}{%
  \bibinfo {author} {\bibfnamefont{M.~A.~H.}\ \bibnamefont{Vozmediano}},
  \bibinfo {author} {\bibfnamefont{M.~I.}\ \bibnamefont{Katsnelson}},\ and\
  \bibinfo {author} {\bibfnamefont{F.}~\bibnamefont{Guinea}},\ }%
  \bibfield{journal}{%
  \Doi{10.1016/j.physrep.2010.07.003}{\bibinfo {journal} {Physics Reports}}\ }%
  \textbf{\bibinfo {volume} {496}},\ \bibinfo {pages} {109} (\bibinfo {year}
  {2010})%
  \bibAnnoteFile{NoStop}{VKG10}%
\bibitem{PC09}%
  \BibitemOpen
  \bibfield{author}{%
  \bibinfo {author} {\bibfnamefont{V.~M.}\ \bibnamefont{Pereira}}\ and\
  \bibinfo {author} {\bibfnamefont{A.~H.}\ \bibnamefont{Castro~Neto}},\ }%
  \bibfield{journal}{%
  \Doi{10.1103/PhysRevLett.103.046801}{\bibinfo {journal} {Phys. Rev. Lett.}}\
  }%
  \textbf{\bibinfo {volume} {103}},\ \bibinfo {pages} {046801} (\bibinfo {year}
  {2009})%
  \bibAnnoteFile{NoStop}{PC09}%
\bibitem{GKG10}%
  \BibitemOpen
  \bibfield{author}{%
  \bibinfo {author} {\bibfnamefont{F.}~\bibnamefont{Guinea}}, \bibinfo {author}
  {\bibfnamefont{M.~I.}\ \bibnamefont{Katsnelson}},\ and\ \bibinfo {author}
  {\bibfnamefont{A.~K.}\ \bibnamefont{Geim}},\ }%
  \bibfield{journal}{%
  \Doi{10.1038/nphys1420}{\bibinfo {journal} {Nat. Phys.}}\ }%
  \textbf{\bibinfo {volume} {6}},\ \bibinfo {pages} {30} (\bibinfo {year}
  {2010})%
  \bibAnnoteFile{NoStop}{GKG10}%
\bibitem{JCV11}%
  \BibitemOpen
  \bibfield{author}{%
  \bibinfo {author} {\bibfnamefont{F.}~\bibnamefont{de~Juan}}, \bibinfo
  {author} {\bibfnamefont{A.}~\bibnamefont{Cortijo}}, \bibinfo {author}
  {\bibfnamefont{M.~A.~H.}\ \bibnamefont{Vozmediano}},\ and\ \bibinfo {author}
  {\bibfnamefont{A.}~\bibnamefont{Cano}},\ }%
  \bibfield{journal}{%
  \Doi{10.1038/nphys2034}{\bibinfo {journal} {Nat. Phys.}}\ }%
  \textbf{\bibinfo {volume} {7}},\ \bibinfo {pages} {810} (\bibinfo {year}
  {2011})%
  \bibAnnoteFile{NoStop}{JCV11}%
\bibitem{LBM10}%
  \BibitemOpen
  \bibfield{author}{%
  \bibinfo {author} {\bibfnamefont{N.}~\bibnamefont{Levy}}, \bibinfo {author}
  {\bibfnamefont{S.~A.}\ \bibnamefont{Burke}}, \bibinfo {author}
  {\bibfnamefont{K.~L.}\ \bibnamefont{Meaker}}, \bibinfo {author}
  {\bibfnamefont{M.}~\bibnamefont{Panlasigui}}, \bibinfo {author}
  {\bibfnamefont{A.}~\bibnamefont{Zettl}}, \bibinfo {author}
  {\bibfnamefont{F.}~\bibnamefont{Guinea}}, \bibinfo {author}
  {\bibfnamefont{A.~H.}\ \bibnamefont{{Castro Neto}}},\ and\ \bibinfo {author}
  {\bibfnamefont{M.~F.}\ \bibnamefont{Crommie}},\ }%
  \bibfield{journal}{%
  \bibinfo {journal} {Science}\ }%
  \textbf{\bibinfo {volume} {329}},\ \bibinfo {pages} {544} (\bibinfo {year}
  {2010})%
  \bibAnnoteFile{NoStop}{LBM10}%
\bibitem{PL09}%
  \BibitemOpen
  \bibfield{author}{%
  \bibinfo {author} {\bibfnamefont{C.-H.}\ \bibnamefont{Park}}\ and\ \bibinfo
  {author} {\bibfnamefont{S.~G.}\ \bibnamefont{Louie}},\ }%
  \bibfield{journal}{%
  \Doi{10.1021/nl803706c}{\bibinfo {journal} {Nano Letters}}\ }%
  \textbf{\bibinfo {volume} {9}},\ \bibinfo {pages} {1793} (\bibinfo {year}
  {2009})%
  \bibAnnoteFile{NoStop}{PL09}%
\bibitem{GSP09}%
  \BibitemOpen
  \bibfield{author}{%
  \bibinfo {author} {\bibfnamefont{M.}~\bibnamefont{Gibertini}}, \bibinfo
  {author} {\bibfnamefont{A.}~\bibnamefont{Singha}}, \bibinfo {author}
  {\bibfnamefont{V.}~\bibnamefont{Pellegrini}}, \bibinfo {author}
  {\bibfnamefont{M.}~\bibnamefont{Polini}}, \bibinfo {author}
  {\bibfnamefont{G.}~\bibnamefont{Vignale}}, \bibinfo {author}
  {\bibfnamefont{A.}~\bibnamefont{Pinczuk}}, \bibinfo {author}
  {\bibfnamefont{L.~N.}\ \bibnamefont{Pfeiffer}},\ and\ \bibinfo {author}
  {\bibfnamefont{K.~W.}\ \bibnamefont{West}},\ }%
  \bibfield{journal}{%
  \Doi{10.1103/PhysRevB.79.241406}{\bibinfo {journal} {Phys. Rev. B}}\ }%
  \textbf{\bibinfo {volume} {79}},\ \bibinfo {pages} {241406} (\bibinfo {year}
  {2009})%
  \bibAnnoteFile{NoStop}{GSP09}%
\bibitem{SGK11}%
  \BibitemOpen
  \bibfield{author}{%
  \bibinfo {author} {\bibfnamefont{A.}~\bibnamefont{Singha}}, \bibinfo {author}
  {\bibfnamefont{M.}~\bibnamefont{Gibertini}}, \bibinfo {author}
  {\bibfnamefont{B.}~\bibnamefont{Karmakar}}, \bibinfo {author}
  {\bibfnamefont{S.}~\bibnamefont{Yuan}}, \bibinfo {author}
  {\bibfnamefont{M.}~\bibnamefont{Polini}}, \bibinfo {author}
  {\bibfnamefont{G.}~\bibnamefont{Vignale}}, \bibinfo {author}
  {\bibfnamefont{M.~I.}\ \bibnamefont{Katsnelson}}, \bibinfo {author}
  {\bibfnamefont{A.}~\bibnamefont{Pinczuk}}, \bibinfo {author}
  {\bibfnamefont{L.~N.}\ \bibnamefont{Pfeiffer}}, \bibinfo {author}
  {\bibfnamefont{K.~W.}\ \bibnamefont{West}},\ and\ \bibinfo {author}
  {\bibfnamefont{V.}~\bibnamefont{Pellegrini}},\ }%
  \bibfield{journal}{%
  \Doi{10.1126/science.1204333}{\bibinfo {journal} {Science}}\ }%
  \textbf{\bibinfo {volume} {332}},\ \bibinfo {pages} {1176} (\bibinfo {year}
  {2011})%
  \bibAnnoteFile{NoStop}{SGK11}%
\bibitem{GMK12}%
  \BibitemOpen
  \bibfield{author}{%
  \bibinfo {author} {\bibfnamefont{K.~K.}\ \bibnamefont{Gomes}}, \bibinfo
  {author} {\bibfnamefont{W.}~\bibnamefont{Mar}}, \bibinfo {author}
  {\bibfnamefont{W.}~\bibnamefont{Ko}}, \bibinfo {author}
  {\bibfnamefont{F.}~\bibnamefont{Guinea}},\ and\ \bibinfo {author}
  {\bibfnamefont{H.~C.}\ \bibnamefont{Manoharan}},\ }%
  \bibfield{journal}{%
  \bibinfo {journal} {Nature}\ }%
  \textbf{\bibinfo {volume} {483}},\ \bibinfo {pages} {306} (\bibinfo {year}
  {2012})%
  \bibAnnoteFile{NoStop}{GMK12}%
\bibitem{HCM07}%
  \BibitemOpen
  \bibfield{author}{%
  \bibinfo {author} {\bibfnamefont{C.-Y.}\ \bibnamefont{Hou}}, \bibinfo
  {author} {\bibfnamefont{C.}~\bibnamefont{Chamon}},\ and\ \bibinfo {author}
  {\bibfnamefont{C.}~\bibnamefont{Mudry}},\ }%
  \bibfield{journal}{%
  \Doi{10.1103/PhysRevLett.98.186809}{\bibinfo {journal} {Phys. Rev. Lett.}}\
  }%
  \textbf{\bibinfo {volume} {98}},\ \bibinfo {pages} {186809} (\bibinfo {year}
  {2007})%
  \bibAnnoteFile{NoStop}{HCM07}%
\bibitem{B12}%
  \BibitemOpen
  \bibfield{author}{%
  \bibinfo {author} {\bibfnamefont{D.~L.}\ \bibnamefont{Bergman}},\ }%
  \bibfield{journal}{%
  \bibinfo {journal} {arxiv:1205.4731}}%
   (\bibinfo {year} {2012})%
  \bibAnnoteFile{NoStop}{B12}%
\bibitem{GGH12}%
  \BibitemOpen
  \bibfield{author}{%
  \bibinfo {author} {\bibfnamefont{P.}~\bibnamefont{Ghaemi}}, \bibinfo {author}
  {\bibfnamefont{S.}~\bibnamefont{Gopalakrishnan}},\ and\ \bibinfo {author}
  {\bibfnamefont{T.~L.}\ \bibnamefont{Hughes}},\ }%
  \bibfield{journal}{%
  \bibinfo {journal} {arxiv:1205.4728}}%
   (\bibinfo {year} {2012})%
  \bibAnnoteFile{NoStop}{GGH12}%
\bibitem{GGR12}%
  \BibitemOpen
  \bibfield{author}{%
  \bibinfo {author} {\bibfnamefont{S.}~\bibnamefont{Gopalakrishnan}}, \bibinfo
  {author} {\bibfnamefont{P.}~\bibnamefont{Ghaemi}},\ and\ \bibinfo {author}
  {\bibfnamefont{S.}~\bibnamefont{Ryu}},\ }%
  \bibfield{journal}{%
  \Doi{10.1103/PhysRevB.86.081403}{\bibinfo {journal} {Phys. Rev. B}}\ }%
  \textbf{\bibinfo {volume} {86}},\ \bibinfo {pages} {081403} (\bibinfo {year}
  {2012})%
  \bibAnnoteFile{NoStop}{GGR12}%
\bibitem{GGV93}%
  \BibitemOpen
  \bibfield{author}{%
  \bibinfo {author} {\bibfnamefont{J.}~\bibnamefont{Gonz\'alez}}, \bibinfo
  {author} {\bibfnamefont{F.}~\bibnamefont{Guinea}},\ and\ \bibinfo {author}
  {\bibfnamefont{M.}~\bibnamefont{Vozmediano}},\ }%
  \bibfield{journal}{%
  \Doi{10.1016/0550-3213(93)90009-E}{\bibinfo {journal} {Nuclear Physics B}}\
  }%
  \textbf{\bibinfo {volume} {406}},\ \bibinfo {pages} {771 } (\bibinfo {year}
  {1993})%
  \bibAnnoteFile{NoStop}{GGV93}%
\bibitem{WAF10}%
  \BibitemOpen
  \bibfield{author}{%
  \bibinfo {author} {\bibfnamefont{R.~T.}\ \bibnamefont{Weitz}}, \bibinfo
  {author} {\bibfnamefont{M.~T.}\ \bibnamefont{Allen}}, \bibinfo {author}
  {\bibfnamefont{B.~E.}\ \bibnamefont{Feldman}}, \bibinfo {author}
  {\bibfnamefont{J.}~\bibnamefont{Martin}},\ and\ \bibinfo {author}
  {\bibfnamefont{A.}~\bibnamefont{Yacoby}},\ }%
  \bibfield{journal}{%
  \Doi{10.1126/science.1194988}{\bibinfo {journal} {Science}}\ }%
  \textbf{\bibinfo {volume} {330}},\ \bibinfo {pages} {812} (\bibinfo {year}
  {2010})%
  \bibAnnoteFile{NoStop}{WAF10}%
\bibitem{MEM11}%
  \BibitemOpen
  \bibfield{author}{%
  \bibinfo {author} {\bibfnamefont{A.~S.}\ \bibnamefont{Mayorov}}, \bibinfo
  {author} {\bibfnamefont{D.~C.}\ \bibnamefont{Elias}}, \bibinfo {author}
  {\bibfnamefont{M.}~\bibnamefont{Mucha-Kruczynski}}, \bibinfo {author}
  {\bibfnamefont{R.~V.}\ \bibnamefont{Gorbachev}}, \bibinfo {author}
  {\bibfnamefont{T.}~\bibnamefont{Tudorovskiy}}, \bibinfo {author}
  {\bibfnamefont{A.}~\bibnamefont{Zhukov}}, \bibinfo {author}
  {\bibfnamefont{S.~V.}\ \bibnamefont{Morozov}}, \bibinfo {author}
  {\bibfnamefont{M.~I.}\ \bibnamefont{Katsnelson}}, \bibinfo {author}
  {\bibfnamefont{V.~I.}\ \bibnamefont{Fal’ko}}, \bibinfo {author}
  {\bibfnamefont{A.~K.}\ \bibnamefont{Geim}},\ and\ \bibinfo {author}
  {\bibfnamefont{K.~S.}\ \bibnamefont{Novoselov}},\ }%
  \bibfield{journal}{%
  \Doi{10.1126/science.1208683}{\bibinfo {journal} {Science}}\ }%
  \textbf{\bibinfo {volume} {333}},\ \bibinfo {pages} {860} (\bibinfo {year}
  {2011})%
  \bibAnnoteFile{NoStop}{MEM11}%
\bibitem{VJB12}%
  \BibitemOpen
  \bibfield{author}{%
  \bibinfo {author} {\bibfnamefont{J.}~\bibnamefont{Velasco}}, \bibinfo
  {author} {\bibfnamefont{L.}~\bibnamefont{Jing}}, \bibinfo {author}
  {\bibfnamefont{W.}~\bibnamefont{Bao}}, \bibinfo {author}
  {\bibfnamefont{Y.}~\bibnamefont{Lee}}, \bibinfo {author}
  {\bibfnamefont{P.}~\bibnamefont{Kratz}}, \bibinfo {author}
  {\bibfnamefont{V.}~\bibnamefont{Aji}}, \bibinfo {author}
  {\bibfnamefont{M.}~\bibnamefont{Bockrath}}, \bibinfo {author}
  {\bibfnamefont{C.~N.}\ \bibnamefont{Lau}}, \bibinfo {author}
  {\bibfnamefont{C.}~\bibnamefont{Varma}}, \bibinfo {author}
  {\bibfnamefont{R.}~\bibnamefont{Stillwell}}, \bibinfo {author}
  {\bibfnamefont{D.}~\bibnamefont{Smirnov}}, \bibinfo {author}
  {\bibfnamefont{F.}~\bibnamefont{Zhang}}, \bibinfo {author}
  {\bibfnamefont{J.}~\bibnamefont{Jung}},\ and\ \bibinfo {author}
  {\bibfnamefont{A.~H.}\ \bibnamefont{MacDonald}},\ }%
  \bibfield{journal}{%
  \Doi{10.1038/nnano.2011.251}{\bibinfo {journal} {Nat. Nanotech.}}\ }%
  \textbf{\bibinfo {volume} {7}},\ \bibinfo {pages} {156} (\bibinfo {year}
  {2012})%
  \bibAnnoteFile{NoStop}{VJB12}%
\bibitem{FS93}%
  \BibitemOpen
  \bibfield{author}{%
  \bibinfo {author} {\bibfnamefont{J.}~\bibnamefont{Fr\"ohlich}}\ and\ \bibinfo
  {author} {\bibfnamefont{U.~M.}\ \bibnamefont{Studer}},\ }%
  \bibfield{journal}{%
  \Doi{10.1103/RevModPhys.65.733}{\bibinfo {journal} {Rev. Mod. Phys.}}\ }%
  \textbf{\bibinfo {volume} {65}},\ \bibinfo {pages} {733} (\bibinfo {year}
  {1993})%
  \bibAnnoteFile{NoStop}{FS93}%
\bibitem{T08}%
  \BibitemOpen
  \bibfield{author}{%
  \bibinfo {author} {\bibfnamefont{I.~V.}\ \bibnamefont{Tokatly}},\ }%
  \bibfield{journal}{%
  \Doi{10.1103/PhysRevLett.101.106601}{\bibinfo {journal} {Phys. Rev. Lett.}}\
  }%
  \textbf{\bibinfo {volume} {101}},\ \bibinfo {pages} {106601} (\bibinfo {year}
  {2008})%
  \bibAnnoteFile{NoStop}{T08}%
\bibitem{WZ84}%
  \BibitemOpen
  \bibfield{author}{%
  \bibinfo {author} {\bibfnamefont{F.}~\bibnamefont{Wilczek}}\ and\ \bibinfo
  {author} {\bibfnamefont{A.}~\bibnamefont{Zee}},\ }%
  \bibfield{journal}{%
  \Doi{10.1103/PhysRevLett.52.2111}{\bibinfo {journal} {Phys. Rev. Lett.}}\ }%
  \textbf{\bibinfo {volume} {52}},\ \bibinfo {pages} {2111} (\bibinfo {year}
  {1984})%
  \bibAnnoteFile{NoStop}{WZ84}%
\bibitem{OBS05}%
  \BibitemOpen
  \bibfield{author}{%
  \bibinfo {author} {\bibfnamefont{K.}~\bibnamefont{Osterloh}}, \bibinfo
  {author} {\bibfnamefont{M.}~\bibnamefont{Baig}}, \bibinfo {author}
  {\bibfnamefont{L.}~\bibnamefont{Santos}}, \bibinfo {author}
  {\bibfnamefont{P.}~\bibnamefont{Zoller}},\ and\ \bibinfo {author}
  {\bibfnamefont{M.}~\bibnamefont{Lewenstein}},\ }%
  \bibfield{journal}{%
  \Doi{10.1103/PhysRevLett.95.010403}{\bibinfo {journal} {Phys. Rev. Lett.}}\
  }%
  \textbf{\bibinfo {volume} {95}},\ \bibinfo {pages} {010403} (\bibinfo {year}
  {2005})%
  \bibAnnoteFile{NoStop}{OBS05}%
\bibitem{RJO05}%
  \BibitemOpen
  \bibfield{author}{%
  \bibinfo {author} {\bibfnamefont{J.}~\bibnamefont{Ruseckas}}, \bibinfo
  {author} {\bibfnamefont{G.}~\bibnamefont{Juzeli\ifmmode~\bar{u}\else
  \={u}\fi{}nas}}, \bibinfo {author}
  {\bibfnamefont{P.}~\bibnamefont{\"Ohberg}},\ and\ \bibinfo {author}
  {\bibfnamefont{M.}~\bibnamefont{Fleischhauer}},\ }%
  \bibfield{journal}{%
  \Doi{10.1103/PhysRevLett.95.010404}{\bibinfo {journal} {Phys. Rev. Lett.}}\
  }%
  \textbf{\bibinfo {volume} {95}},\ \bibinfo {pages} {010404} (\bibinfo {year}
  {2005})%
  \bibAnnoteFile{NoStop}{RJO05}%
\bibitem{DGJ11}%
  \BibitemOpen
  \bibfield{author}{%
  \bibinfo {author} {\bibfnamefont{J.}~\bibnamefont{Dalibard}}, \bibinfo
  {author} {\bibfnamefont{F.}~\bibnamefont{Gerbier}}, \bibinfo {author}
  {\bibfnamefont{G.}~\bibnamefont{Juzeli\ifmmode~\bar{u}\else \={u}\fi{}nas}},\
  and\ \bibinfo {author} {\bibfnamefont{P.}~\bibnamefont{\"Ohberg}},\ }%
  \bibfield{journal}{%
  \Doi{10.1103/RevModPhys.83.1523}{\bibinfo {journal} {Rev. Mod. Phys.}}\ }%
  \textbf{\bibinfo {volume} {83}},\ \bibinfo {pages} {1523} (\bibinfo {year}
  {2011})%
  \bibAnnoteFile{NoStop}{DGJ11}%
\bibitem{SGG12}%
  \BibitemOpen
  \bibfield{author}{%
  \bibinfo {author} {\bibfnamefont{P.}~\bibnamefont{San-Jose}}, \bibinfo
  {author} {\bibfnamefont{J.}~\bibnamefont{Gonz\'alez}},\ and\ \bibinfo
  {author} {\bibfnamefont{F.}~\bibnamefont{Guinea}},\ }%
  \bibfield{journal}{%
  \Doi{10.1103/PhysRevLett.108.216802}{\bibinfo {journal} {Phys. Rev. Lett.}}\
  }%
  \textbf{\bibinfo {volume} {108}},\ \bibinfo {pages} {216802} (\bibinfo {year}
  {2012})%
  \bibAnnoteFile{NoStop}{SGG12}%
\bibitem{WY75}%
  \BibitemOpen
  \bibfield{author}{%
  \bibinfo {author} {\bibfnamefont{T.~T.}\ \bibnamefont{Wu}}\ and\ \bibinfo
  {author} {\bibfnamefont{C.~N.}\ \bibnamefont{Yang}},\ }%
  \bibfield{journal}{%
  \Doi{10.1103/PhysRevD.12.3843}{\bibinfo {journal} {Phys. Rev. D}}\ }%
  \textbf{\bibinfo {volume} {12}},\ \bibinfo {pages} {3843} (\bibinfo {year}
  {1975})%
  \bibAnnoteFile{NoStop}{WY75}%
\bibitem{FSW97}%
  \BibitemOpen
  \bibfield{author}{%
  \bibinfo {author} {\bibfnamefont{M.~H.}\ \bibnamefont{Friedman}}, \bibinfo
  {author} {\bibfnamefont{Y.}~\bibnamefont{Srivastava}},\ and\ \bibinfo
  {author} {\bibfnamefont{A.}~\bibnamefont{Widom}},\ }%
  \bibfield{journal}{%
  \bibinfo {journal} {J. Phys. G: Nucl. Part. Phys.}\ }%
  \textbf{\bibinfo {volume} {23}},\ \bibinfo {pages} {1061} (\bibinfo {year}
  {1997})%
  \bibAnnoteFile{NoStop}{FSW97}%
\bibitem{Note1}%
  \BibitemOpen
  \bibinfo {note} {The resolution to this apparent paradox is simply that for
  non-abelian fields there are further independent gauge invariant quantities,
  such as $D_iF_{jk}$, that distinguish among them.}%
  \bibAnnoteFile{Stop}{Note1}%
\bibitem{B08}%
  \BibitemOpen
  \bibfield{author}{%
  \bibinfo {author} {\bibfnamefont{D.~M.}\ \bibnamefont{Basko}},\ }%
  \bibfield{journal}{%
  \Doi{10.1103/PhysRevB.78.125418}{\bibinfo {journal} {Phys. Rev. B}}\ }%
  \textbf{\bibinfo {volume} {78}},\ \bibinfo {pages} {125418} (\bibinfo {month}
  {Sep}\ \bibinfo {year} {2008})%
  \bibAnnoteFile{NoStop}{B08}%
\bibitem{M07}%
  \BibitemOpen
  \bibfield{author}{%
  \bibinfo {author} {\bibfnamefont{J.~L.}\ \bibnamefont{Ma\~nes}},\ }%
  \bibfield{journal}{%
  \Doi{10.1103/PhysRevB.76.045430}{\bibinfo {journal} {Phys. Rev. B}}\ }%
  \textbf{\bibinfo {volume} {76}},\ \bibinfo {pages} {045430} (\bibinfo {year}
  {2007})%
  \bibAnnoteFile{NoStop}{M07}%
\bibitem{WZ10}%
  \BibitemOpen
  \bibfield{author}{%
  \bibinfo {author} {\bibfnamefont{R.}~\bibnamefont{Winkler}}\ and\ \bibinfo
  {author} {\bibfnamefont{U.}~\bibnamefont{Z\"ulicke}},\ }%
  \bibfield{journal}{%
  \Doi{10.1103/PhysRevB.82.245313}{\bibinfo {journal} {Phys. Rev. B}}\ }%
  \textbf{\bibinfo {volume} {82}},\ \bibinfo {pages} {245313} (\bibinfo {year}
  {2010})%
  \bibAnnoteFile{NoStop}{WZ10}%
\bibitem{L12}%
  \BibitemOpen
  \bibfield{author}{%
  \bibinfo {author} {\bibfnamefont{T.~L.}\ \bibnamefont{Linnik}},\ }%
  \bibfield{journal}{%
  \bibinfo {journal} {J. Phys.: Cond. Mat.}\ }%
  \textbf{\bibinfo {volume} {24}},\ \bibinfo {pages} {205302} (\bibinfo {year}
  {2012})%
  \bibAnnoteFile{NoStop}{L12}%
\bibitem{Falko}%
  \BibitemOpen
  \bibfield{author}{%
  \bibinfo {author} {\bibfnamefont{R.}~\bibnamefont{Ferone}}, \bibinfo {author}
  {\bibfnamefont{J.~R.}\ \bibnamefont{Wallbank}}, \bibinfo {author}
  {\bibfnamefont{V.}~\bibnamefont{Z\'olyomi}}, \bibinfo {author}
  {\bibfnamefont{E.}~\bibnamefont{McCann}},\ and\ \bibinfo {author}
  {\bibfnamefont{V.~I.}\ \bibnamefont{Fal’ko}},\ }%
  \bibfield{journal}{%
  \Doi{10.1016/j.ssc.2011.05.017}{\bibinfo {journal} {Solid State Comm.}}\ }%
  \textbf{\bibinfo {volume} {151}},\ \bibinfo {pages} {1071 } (\bibinfo {year}
  {2011})%
  \bibAnnoteFile{NoStop}{Falko}%
\bibitem{Pankratov}%
  \BibitemOpen
  \bibfield{author}{%
  \bibinfo {author} {\bibfnamefont{O.}~\bibnamefont{Pankratov}}, \bibinfo
  {author} {\bibfnamefont{S.}~\bibnamefont{Hensel}},\ and\ \bibinfo {author}
  {\bibfnamefont{M.}~\bibnamefont{Bockstedte}},\ }%
  \bibfield{journal}{%
  \Doi{10.1103/PhysRevB.82.121416}{\bibinfo {journal} {Phys. Rev. B}}\ }%
  \textbf{\bibinfo {volume} {82}},\ \bibinfo {pages} {121416} (\bibinfo {month}
  {Sep}\ \bibinfo {year} {2010})%
  \bibAnnoteFile{NoStop}{Pankratov}%
\bibitem{Note2}%
  \BibitemOpen
  \bibinfo {note} {Or to the Dirac Hamiltonian in the presence of Rashba
  spin-orbit coupling. Not surprisingly these Hamiltonians are described in
  terms of SU(2) gauge fields too.}%
  \bibAnnoteFile{Stop}{Note2}%
\bibitem{GHS98}%
  \BibitemOpen
  \bibfield{author}{%
  \bibinfo {author} {\bibfnamefont{V.~P.}\ \bibnamefont{Gusynin}}, \bibinfo
  {author} {\bibfnamefont{D.~K.}\ \bibnamefont{Hong}},\ and\ \bibinfo {author}
  {\bibfnamefont{I.~A.}\ \bibnamefont{Shovkovy}},\ }%
  \bibfield{journal}{%
  \Doi{10.1103/PhysRevD.57.5230}{\bibinfo {journal} {Phys. Rev. D}}\ }%
  \textbf{\bibinfo {volume} {57}},\ \bibinfo {pages} {5230} (\bibinfo {year}
  {1998})%
  \bibAnnoteFile{NoStop}{GHS98}%
\end{thebibliography}%

\end{document}